\documentclass[twocolumn]{article}
\usepackage{amsfonts,amssymb,amsmath,amscd}
\usepackage{hyperref}
\usepackage{xurl}

\title{
Using Generative AI for Literature Searches and Scholarly Writing:\\
Is the Integrity of the Scientific Discourse in Jeopardy?\,\footnote{Version of September 1, 2023. Accepted for publication in \!{\em Notices of the AMS}.}
}

\author{
Paul G.~Schmidt\,\footnote{Professor Emeritus, Department of Mathematics and Statistics, Auburn University, \url{mailto:schmipg@auburn.edu}}
\and
Amnon J.~Meir\,\footnote{Professor, Department of Mathematics, Southern Method\-ist University, \url{mailto:ajmeir@smu.edu}}
}

\date{}

\begin{document}

\maketitle

\abstract{
Ever since the public release of ChatGPT in November  2022, serious concerns have been raised about the impact and potentially dire consequences of the increasingly widespread use of generative AI tools for purposes of scientific writing and publishing. We document the ongoing discussion in the science community with a review of news articles, editorials, and position statements by major scientific publishers; discuss the potential pitfalls of using generative AI tools such as ChatGPT as aids in scholarly writing; furnish evidence for the proposition that AI-induced contamination of the scientific literature is not only a threat, but already a reality; and call upon leaders in our field to develop policies and guidelines to stem the spread of such contamination. Closing on a positive note, we provide a brief overview of potentially useful capabilities and sensible applications of ChatGPT and similar AI tools for purposes of scholarly writing.
}

\section{Introduction}\label{Intro}

With the public release of ChatGPT on November~30, 2022, AI has gone mainstream. We are aware of colleagues who are using ChatGPT or other generative AI tools to assist with literature searches and the writing of research papers, proposals, and reviews. The practice is likely to spread rapidly. While the advent of broadly available generative AI tools like ChatGPT has spawned exciting new possibilities in numerous fields, including scholarly writing, there is also a huge potential for abuse. Ever since the public release of ChatGPT, serious concerns have been raised about the impact and potentially dire consequences of the widespread use of generative AI tools in scientific writing and publishing.$^\text{17-21}$\,\footnote[4]{Since {\em Notices of the AMS\/} does not allow extensive bibliographies, most of the citations in this article, marked by numerical superscripts, refer to the bibliography of~\cite{Supplement}, an online collection of supplementary material for this article, available at SMU Scholar. The bibliography of [1] contains {\em all} our references, including the sixteen listed in this article (with the exception of [1]), the ones cited but not listed in this article, and additional titles cited only in~\cite{Supplement}.}

In a recent issue of the journal {\em ACS Energy Letters} \cite{Grimaldi}, published by the American Chemical Society (ACS), Grimaldi and Ehrler write that ``a text-generation system combining speed of implementation with eloquent and structured language could enable a leap forward for the serialized production of scientific-looking papers devoid of scientific content, increasing the throughput of paper factories and making detection of fake research more time-consuming.'' Observing that ChatGPT and other generative AI tools can write papers that might well pass peer review for a perspective article, the authors point out that ``this ability makes the urgent need for a code of conduct for the use of AI-generated text in scientific literature abundantly clear.''

Within a month after the public release of ChatGPT,  the editors of flagship journals like {\em Science} and {\em Nature}, numerous publishers of scientific literature, and the boards of e-print repositories such as arXiv, ChemRxiv, bioRxiv, and medRxiv were discussing policies and guidelines regarding the use of AI in scholarly writing~\cite{Stokel-Walker}. On January 26, 2023, {\em Science} published an editorial by editor-in-chief H.~Holden Thorp~\cite{Thorp}, explaining the rationale behind a recent change to the editorial policies for the {\em Science} group of journals~\cite{Science}: ``Text generated from AI, machine learning, or similar algorithmic tools cannot be used in papers published in Science journals, nor can the accompanying figures, images, or graphics be the products of such tools, without explicit permission from the editors. In addition, an AI program cannot be an author of a Science journal paper. A violation of this policy constitutes scientific misconduct.''

On January 24, 2023, {\em Nature} published an editorial~\cite{NatureEditorial}, postulating that ``tools such as ChatGPT threaten transparent science'' and suggesting that ``researchers should ask themselves how the transparency and trust-worthiness that the process of generating knowledge relies on can be maintained if they or their colleagues use software that works in a fundamentally opaque manner.''

To address the threat, {\em Nature}, along with all the Springer Nature journals, adopted the following ground rules~\cite{NatureEditorial} for the use of large language models (LLMs): ``First, no LLM tool will be accepted as a credited author on a research paper. That is because any attribution of authorship carries with it accountability for the work, and AI tools cannot take such responsibility. Second, researchers using LLM tools should document this use in the methods or acknowledgements sections. If a paper does not include these sections, the introduction or another appropriate section can be used to document the use of the LLM.'' Note that {\em Nature}'s policy on AI use in authoring is much less encompassing and much more permissive than {\em Science}'s.

On February 13, 2023, the Committee for Publication Ethics (COPE) published a position statement on AI and authorship \cite{COPE-Position}, closely mirroring the two principles formulated in the {\em Nature} editorial. Numerous editors, publishers, research institutes, corporations, universities, and professional organizations across all fields of academia are members of COPE and adhere to its code of conduct;$^{22}$ current membership exceeds 14,000 entities.$^{23}$ Many major publishers have subsequently issued detailed guidelines for authors, reviewers, and editors governing the use of generative AI tools.$^\text{24-28}$

Different from the American Physical Society (APS), the American Chemical Society (ACS), and many other professional societies in the sciences, neither SIAM nor AMS are members of COPE. As far as we are aware, and as of this writing, SIAM has not published any policy or guidelines regarding the use of AI tools in scholarly writing. We should know, as one of us serves on the editorial board of the {\em SIAM Journal on Numerical Analysis} (SINUM).

A SIAM web page entitled ``Information for Journal Authors''$^{29}$ includes a section on ``Authorial Integrity in Scientific Publication,'' at the end of which a subsection on ``Other Resources'' contains a link to the COPE homepage,$^{30}$ where the diligent reader may find COPE's position statement on AI use in scholarly writing~\cite{COPE-Position}. Another link, unfortunately broken, should lead to the whitepaper ``Recommendations for Promoting Integrity in Scientific Journal Publications,''$^{31}$ issued by the Council of Scientific Editors (CSE). Section~2.1.15 and part of Section~2.2.2 of the pdf version of this white paper address the use of AI in scholarly writing. The relevant paragraphs were added on April 25, 2023, and are still missing in the html version of the whitepaper. On June~22, 2023, CSE hosted a webinar on ``Updates to CSE's Editorial Policy Recommendations'';$^{32}$ the first item on the agenda was to ``explore the background of the recent guidance on artificial intelligence and chatbots and how it affects scholarly publishing.''

Only recently, some time this past June, the AMS amended its ``Ethical Guidelines and Journal Policies''~\cite{AMS} with a section on the use of artificial intelligence in authoring: ``The following statement, issued by the Committee on Publication Ethics in February 2023, has been adopted by the AMS Committee on Publications.'' What follows is the three-paragraph COPE position statement referred to earlier~\cite{COPE-Position}, declaring that AI tools cannot be authors and that authors using AI tools must fully disclose any such usage.

Adopting the COPE position statement appears to be little more than a stop-gap measure, pending the drafting of more specific AMS policies and guidelines for authors, referees, and editors. In fact, AMS has recently established an ``Advisory Group on Artificial Intelligence and the Mathematical Community,'' authorized by AMS president Bryna Kra on July 10, 2023, which is charged with discussing ``the role of mathematics in the development and deployment of artificial intelligence, the impact of artificial intelligence on research in mathematics, the use of AI in publications, education, and research, and the impact of AI on our community.''\cite{AMS-Advisory}

We are fairly certain that also SIAM and other professional organizations in our field have, by now, formed advisory committees charged with the drafting of policies and guidelines regarding the use of generative AI in scientific writing, reviewing, and publishing; yet they appear to be biding their time, while time is running out.

Professional societies in other fields are much further along. The ``Editorial Policies and Practices'' of journals published by the American Physical Society (APS) contain a statement on the ``Appropriate Use of AI-Based Writing Tools'' that addresses both authors and referees.$^{33}$ Already on February~27, 2023, the American Chemical Society (ACS) issued ``best practices for using AI when writing scientific manuscripts,''$^{34}$ including recommendations for scientists using AI tools at any stage of their research and, in particular, when drafting manuscripts. On April~20, 2023, the Publications Board of the Association for Computing Machinery (ACM) approved a new ``Policy on Authorship''$^{35}$ that includes guidance for the use of generative AI tools. On May~31, 2023, the World Association of Medical Editors (WAME) published ``Recommendations on Chatbots and Generative Artificial Intelligence in Relation to Scholarly Publications.''$^{36}$ Similar recommendations were issued by the International Committee of Medical Journal Editors (ICMJE).$^{37}$ Also the American Medical Association (AMA) has published clear rules for the use of AI in manuscripts submitted to its journals.$^{38}$ All these policies are in line with and expand upon the COPE position~\cite{COPE-Position}.

Also research funding agencies are slowly catching up. In a notice~\cite{NIH} released on June 23, 2023, the Office of Extramural Research of the National Institutes of Health (NIH) informed the community that ``the NIH prohibits NIH scientific peer reviewers from using natural language processors, large language models, or other generative Artificial Intelligence (AI) technologies for analyzing and formulating peer review critiques for grant applications and R\,\&\,D contract proposals.'' The rationale: ``Uploading or sharing content or original concepts from an NIH grant application, contract proposal, or critique to online generative AI tools violates the NIH peer review confidentiality and integrity requirements.''

In a policy paper \cite{ARC} published on July 7, 2023, the Australian Research Council (ARC), a peer institution of the National Science Foundation (NSF) in Australia, banned the use of AI tools by grant assessors (i.e., grant proposal reviewers), postulating that ``the use of generative AI may compromise the integrity of the ARC's peer review process by, for example, producing text that contains inappropriate content, such as generic comments and restatements of the application.'' Moreover, ``when information is entered into generative AI tools, it enters the public domain and can be accessed by unspecified third parties.'' Thus, the ``release of material into generative AI tools constitutes a breach of confidentiality,'' and ARC peer reviewers ``must not use generative AI as part of their assessment activities.''

Remarkably, the ARC policy goes well beyond advice and guidance: ``In cases where the use of generative AI by assessors is suspected, the ARC will remove that assessment from its assessment process. If, following an investigation, an assessor is found to have breached the Code during ARC assessment, the ARC may impose consequential actions in addition to any imposed by the employing institution.'' Here ``the Code'' refers to the Australian Code for the Responsible Conduct of Research. As reported by Research Professional News,$^{39}$ at least one suspect assessor report was actually removed from the peer review process, after a complaint had been lodged. Quoting the Twitter account ARC Tracker,$^{40}$ an article in Times Higher Education$^{41}$ recounts that multiple applicants to the ARC Discovery Projects program have publicly complained that assessor reports they received were ``generic regurgitations of their applications with little evidence of critique, insight or assessment'' and appeared to be written by AI or, specifically, ChatGPT.

Different from its peer institutions NIH and ARC, the National Science Foundation (NSF) has been slow in responding to a development widely recognized as having a profound impact on scholarly activity. The 203-page Proposal and Award Policies and Procedures Guide (PAPPG),$^{42}$ effective January 30, 2023, contains the string ``ai'' only as part of words like email, training, or failure; ``AI'' appears only in RAISE (Research Advanced by Interdisciplinary Science and Engineering); the string ``A.I.'' does not occur at all, and neither does ``Artificial Intelligence.'' The same holds for the ``Instructions for Proposal Review'' found on Fastlane,$^{43}$ which were last modified in September 2017. At the time of this writing, a search of the NSF and Fastlane webpages for policies or guidelines regarding AI use in proposal preparation, proposal review, and progress reports does not yield anything relevant. Given that AI research is a major focus of current NSF programs,$^{44,45}$ we find this surprising, to say the least.

Our objective in writing this article is threefold: to call attention to the fact that AI-induced contamination of the scientific literature is not only a threat, but already a reality; to alert the community to the potential pitfalls of using generative AI tools as aids in scholarly writing; and to call upon leaders in our field to issue adequate policies and guidelines as soon as possible. In fact, this article started out as an open letter that we sent to the leadership of SIAM, AMS, and NSF-MPS. We thank AMS president Bryna Kra for her suggestion to turn the letter into a paper suitable for publication in the {\em Notices of the AMS}. We are grateful to several colleagues, who saw the original letter or early versions of the paper, for their feedback and encouragement. Guillaume Cabanac, who has been at the frontline of the battle against AI-induced contamination of the scientific literature for several years, was instrumental in opening our eyes and helping us uncover evidence of existing contamination. Finally, we thank the anonymous referees for their constructive criticism and helpful suggestions.

The remainder of this article is organized as follows. In Section~\ref{Generative} we present evidence supporting our claim that contamination of the scientific literature by way of generative AI tools is occurring on a surprisingly large scale, and has been occurring for some time. For a much more detailed discussion, quoting chapter and verse, we refer the reader to the supplementary material available online, at SMU Scholar~\cite{Supplement}. In Section~\ref{Experiments} we describe our own experiments with ChatGPT and the concerns they triggered. We show how the na\"ive use of ChatGPT and other generative AI tools may pollute the literature with pseudo-science and potentially serious yet hard to detect falsities and fabrications. Our observations are based on two long conversations with ChatGPT, summaries and verbatim transcripts of which are provided in~\cite{Supplement}. Finally, in Section~\ref{Sensible} we complement our largely negative assessment of ChatGPT's potential as a research assistant with a discussion of some useful features and sensible applications of the tool in scholarly writing.

\section{Generative AI and Scholarship: a Reality Check}\label{Generative}

ChatGPT was publicly released on November 30, 2022. By the end of January 2023, it had reached 100 million active users, making it the fastest-growing consumer software application in history. TikTok needed about nine months after its global launch to achieve this feat, and Instagram 2$\frac12$ years.$^{46}$

ChatGPT is a marvel of generative artificial intelligence, and its successor ChatGPT Plus, powered by OpenAI's latest and most advanced large language model GPT-4 and available by subscription, is even more versatile. Yet perfect they are not. As amply documented in news reports \cite{Weise} and technical papers \cite{OpenAI}, they continue to be plagued by ``hallucinations,'' a concept introduced by Google AI researchers several years ago. Technically, the term ``hallucination'' refers to semantically or syntactically plausible, yet factually incorrect or nonsensical output.$^{47}$ Opinions are split, among AI experts, as to the severity and permanence of this phenomenon.$^\text{48-50}$ In any case, it is a fact that ChatGPT and even today's most advanced chatbots routinely fabricate content and present this fictitious content authoritatively and convincingly. The following are quotes from Open\-AI's ``GPT-4 System Card'' \cite[pp.~59/60]{OpenAI}:

``Despite GPT-4's capabilities, it maintains a tendency to make up facts, to double-down on incorrect information, and to perform tasks incorrectly. Further, it often exhibits these tendencies in ways that are more convincing and believable than earlier GPT models (e.g., due to authoritative tone or to being presented in the context of highly detailed information that is accurate), increasing the risk of overreliance.''

``Overreliance occurs when users excessively trust and depend on the model, potentially leading to unnoticed mistakes and inadequate oversight.''

``As mistakes become harder for the average human user to detect and general trust in the model grows, users are less likely to challenge or verify the model's responses.''

Nota bene: the creators of ChatGPT are explicitly warning us not to trust the tool in any circumstance where veracity is of the essence.$^{51}$

Will such warnings deter people from employing chatbots for ``truth-sensitive'' tasks? Will such warnings at least instill a sense of caution? --- A New York lawyer recently filed a ten-page brief in Federal District Court, citing more than half a dozen judicial precedents relevant to the case he was presenting.$^{52}$ The brief was complete with summaries of court decisions, including direct quotes and internal citations, naming courts and judges, listing dates and docket numbers. Alas, neither the opposing lawyers nor the presiding judge were able to verify these citations. As it turns out, ChatGPT had done the legal research and made it all up. ``I did not comprehend that ChatGPT could fabricate cases,'' the lawyer told the judge in a subsequent hearing.$^{53}$

This is what happens when human stupidity meets artificial intelligence. Scientists, of course, and mathematicians in particular, are immune to the disease. Or are they? --- In the introduction we reported on the case of a peer review of a research proposal submitted to the Australian Research Council (ARC) that was rejected by the agency after the applicant complained that it was clearly written by ChatGPT.$^{39,41}$ The smoking gun? --- Apparently, the review contained, out of place and devoid of meaning, the phrase ``regenerate response.'' This is the label of a button below ChatGPT's output window. If you copy a response issued by ChatGPT, and you do so in a sufficiently sloppy way, you will capture the label as part of the response. If you then paste the response into a piece of your own writing, and you do so without any editing or even reading, the telltale phrase will end up in your paper or review.

Of course, no professional scientist or mathematician would ever fall into this trap --- the case of the ARC referee is surely an anomaly. Maybe they were under a lot of pressure and needed to complete the job really fast. It happened, but it will likely never happen again. Or will it?

PubPeer$^{54}$ is an online platform for the post-publication peer review of scientific papers, operated by the PubPeer Foundation, a California-registered public-benefit corporation with nonprofit status. Since May 2023, at least 28 papers, published in supposedly reputable and peer-reviewed journals, or posted on the arXiv or other e-print repositories, were flagged on PubPeer for the out-of-place occurrence of the phrase ``regenerate response''; at least seven more were flagged for containing the phrase ``as an AI language model,'' which ChatGPT frequently includes in its responses (usually as part of a caveat, warning the user not to blindly trust its output). At the time of this writing, additional papers are being flagged on an almost daily basis.
 
We decided to refrain from directly quoting any of these problematic papers in the present article. However, we furnish a detailed analysis of a representative sample, quoting chapter and verse, in the supplementary material available online, at SMU Scholar~\cite{Supplement}.

One particularly striking example came to our attention just a few days ago: a paper presented at a May 2023 conference on emerging technologies and published in the {\em IEEE Xplore} digital library by the Institute of Electrical and Electronics Engineers (IEEE).$^{55}$ The paper was flagged on PubPeer in July, for including the unexpected phrase ``regenerate response'' at the end of the abstract. The introduction of the paper describes related prior work, referencing Saponas et al. (2009), Wang et al. (2015), Jin et al. (2016), Kang et al. (2019), and Moussa et al. (2020), albeit without providing any further bibliographical details. In early August, a commenter on PubPeer noted that none of these five ``references'' appears in the bibliography at the end of the paper and that all of them are very likely ChatGPT hallucinations. We searched both Scopus and Dimensions~AI for 2009 publications with first author Saponas (a Microsoft researcher) and found two; however, neither one matches the description in the introduction of the suspect paper. The other four ``references'' are harder to unmask, since the names of the first authors are very common; but we are convinced they are hallucinations as well.

In fact, we believe that besides the abstract, also the entire introduction of the paper is AI-generated. The style is vintage ChatGPT, including the vague and imprecise way of describing and referencing prior work --- we have seen this many times in our own experiments. Had the human authors prodded ChatGPT to provide additional bibliographical information about those five ``references,'' it would have happily obliged, producing precise and complete, if completely bogus, references. Too bad that this thought did not occur to them. Most likely, they didn't even notice that ChatGPT's introduction included incomplete references not included in their bibliography\ldots

A number of AI tools are available to detect artificially generated text. One of them is OpenAI's \text{GPT-2} Output Detector, which estimates the probability of a given text being AI-generated. The source code is publicly available;$^{56}$ an easy-to-use implementation can be found online.$^{57}$ According to Open\-AI, the tool ``is able to detect 1.5 billion parameter GPT-2-generated text with approximately 95\% accuracy.''$^{\text{58,\,p.13}}$ The phrase ``1.5 billion parameter GPT-2-generated text'' refers to output produced by the largest GPT-2 model; the claim about the detector tool's accuracy is cryptic at best, and we have been unable to obtain more precise information. However, according to Cabanac et al.~\cite{Cabanac2}, and confirmed by our own experiments, the tool is quite effective in detecting not only \text{GPT-2} output, but also other AI-generated content. As a caveat, we note that AI detection tools have recently come under a lot of fire.$^{59,60}$ In January 2023, Open\-AI launched a new detection tool, which it promptly retracted less than six months later, ``due to its low rate of accuracy.''$^{61,62}$

For whatever it's worth, we fed both the abstract and the introduction of the afore-mentioned paper into the GPT-2 output detector. Both garnered whopping scores of 99.98\% and 99.94\%, respectively. Again, the detector score represents the probability of the respective text being AI-generated. Interestingly, if the unexpected phrase ``regenerate response'' is included, the abstract's detector score is merely 11.71\% --- apparently the detector considers this phrase so weird and out of place that it must be human-generated\ldots

The papers referred to above all feature telltale signs of ChatGPT's unacknowledged involvement in the writing, which typically constitutes a violation of the publisher's code of conduct, but does not necessarily qualify as ``contamination of the scientific literature.'' While the detection of AI-generated content can be automated, parsing suspect content for evidence of plagiarism, factual errors, fabricated reviews, and fictitious references still takes a human expert.

Typically, issues of this kind are uncovered by people who are directly affected, like the afore-mentioned ARC grant applicants who received AI-generated referee reports.$^{39,41}$ Another example, recently reported in Times Higher Education,$^{63}$ is the case of Robin Bauwens, a social scientist at Tilburg University in the Netherlands, whose submission to an Emerald Publishing journal was rejected by a referee. In the report, the referee suggested that Bauwens familiarize himself with several relevant literature reviews in his field, all allegedly authored by Dutch academics unbeknownst to him. References and authors turned out to be fictitious, most likely fabricated by ChatGPT.

Emerald Publishing told Times Higher Education$^{63}$ that it had recently updated its guidelines for authors and referees, stating in particular that ``ChatGPT and other AI tools should not be utilized by reviewers of papers submitted to journals published by Emerald'' and that ``AI tools/LLMs should not replace the peer review process that relies on human subject matter expertise and critical appraisal.'' The mere thought that the latter should warrant emphasizing is cringe-inducing, at least to some of us\ldots

Nobody outside the corporate headquarters of OpenAI knows exactly what data ChatGPT was trained on. There is anecdotal evidence that the MEDLINE database of references and abstracts on life sciences and biomedical topics was included in its training data. In fact, ChatGPT did fairly well in a recent study~\cite{Athaluri} published in the journal {\em Cureus} (part of Springer Nature Group). The experiment started with the following prompt: ``Suggest 50 novel medical research topics that can be performed by undergraduate medical students in India. The topics must be feasible, interesting, novel, ethical, and relevant'' (the so-called FINER criteria). ChatGPT readily obliged and was subsequently instructed to write an elaborate research protocol for each of the 50 topics, including a proper introduction, sections on objectives, methodology, and implications, and a list of references. The protocols were then evaluated by a panel of experts and generally found to be feasible.

However, of the 178 references provided by ChatGPT, 69 did not have a valid DOI, while 28 could not be located at all and are presumed to be fictitious. The authors conclude that ``researchers using ChatGPT should exercise caution in relying solely on the references generated by the AI chatbot.'' We promise to keep that in mind, especially since the reference hallucination rate we observed in our own experiments (see Section~\ref{Experiments}) was much closer to 100\% than the measly 28/178 $\approx$ 16\% found in the {\em Cureus} study.

AI-induced contamination of the scientific literature predates the public release of ChatGPT by several years. In~\cite{Cabanac1} the computer scientists Guillaume Cabanac and Cyril Labb\'e discussed the prevalence in the scientific literature of ``research papers'' devoid of meaningful results and sometimes plainly nonsensical. They suspected that such papers are algorithmically generated by programs like SCIgen or MATHgen and proposed methods of detection and elimination.

In~\cite{Cabanac2} Cabanac, Labb\'e, and Magazinov coined the term {\em tortured phrases}: unexpected strange phrases in lieu of established ones, such as ``counterfeit consciousness'' in lieu of ``artificial intelligence.'' A typical tortured phrase emerges when word-by-word synonymic substitution is applied to a multi-word technical phrase. In this way, ``high-performance computing'' may morph into ``elite figuring,''  ``data warehouse'' into ``information stockroom,'' ``network attack'' into ``organization ambush.'' Nontechnical articles summarizing and discussing the results of \cite{Cabanac1, Cabanac2} appeared in {\em Nature\/}$^{64,65}$ and the Bulletin of the Atomic Scientists.$^{66}$

Cabanac et al.\ made a list of 30 tortured phrases they had encountered in the scientific literature, most involving computer-science parlance. Using the Dimensions AI academic search engine, they exposed more than 800 papers that included at least one of the phrases; 31 of those appeared in a single journal published by Elsevier, {\em Microprocessors and Microsystems}. ``Preliminary probes show that several thousands of papers with tortured phrases are indexed in major databases,'' they write, pointing out that their study focussed on tortured phrases in Computer Science and that ``tortured phrases related to the concepts of other scientific fields are yet to be exposed.''

Text containing tortured phrases typically scores high when run through an AI output detection tool, which estimates the probability of the text being AI-generated. Cabanac et al.~\cite{Cabanac2} used OpenAI's \text{GPT-2} Output Detector to analyze the abstracts of tens of thousands of papers across the scientific literature. They retrieved the abstracts of all articles published in Volumes 80--83 (2019--2021) of {\em Microprocessors and Microsystems} (MPMS) that were processed in less than 30 days (from date received to date accepted), a total of 389 articles. This set of abstracts (Set~1) was tested against several control sets, including the abstracts of 50 articles recently published in MPMS with processing times exceeding 40 days (Set~2), the abstracts of 50 computer-science related articles recently published in SIAM journals (Set~3), and the abstracts of a sample of 139,236 articles published in 2021 by Elsevier (Set~4).

Of the abstracts in Set 1 (MPMS with fast processing), 81\% scored at least 0.9 (90\% probability of being AI-generated), while only 8.5\% scored below 0.1 (10\% probability of being AI-generated). Of the abstracts in Set 2 (MPMS with not so fast processing), 16\% scored at least 0.9, while 78\% scored below 0.1. Of the abstracts in Set 3 (SIAM journals), none scored above 0.7, while 90\% scored below 0.1. Of the abstracts in Set 4 (Elsevier 2021), 3\% scored at least 0.9, while 89.9\% scored below 0.1.

Elsevier journals that published at least 25 articles in 2021 with GPT-2 output detector scores of at least 0.7 include the {\em Journal of Computational and Applied Mathematics} (39 articles, representing 13.6\% of all articles tested), the {\em Journal of Differential Equations} (27, 11.2\%), and the {\em Journal of Mathematical Analysis and Applications} (25, 7.5\%).

Let this sink in for a moment: the {\em Journal of Differential Equations}, the gold standard in the field, published at least 27 articles in 2021 whose abstracts were, with a probability of at least 70\%, AI-generated. We take comfort in the thought that Joseph LaSalle and Jack Hale did not live to witness this moment\ldots

Since the study by Cabanac et al.\ predates the public release of ChatGPT by more than a year, the culprits are earlier generative AI tools. \text{GPT-2} was released in 2019, \text{GPT-3} in 2020; but more likely, the authors of the suspect abstracts employed automatic translation or paraphrasing tools such as SpinBot,$^{67}$ ``a free, automatic article spinner that will rewrite human readable text into additional, intelligent, readable text'' that has been around for more than a decade. Authors resorting to automatic translation may be doing so with the perfectly honest objective of improving their writing in English. Authors employing paraphrasing apps such as SpinBot most likely do so to conceal plagiarism.

As noted before, the mere fact that a research paper was written with the aid of an AI tool does not render it fraudulent or otherwise objectionable, at least not if the fact is acknowledged (which it usually and unfortunately isn't). Obviously, Cabanac et al.\ were not able to fact-check every one of the suspect papers they discovered; they did, however, perform a detailed analysis of a sample of the papers in MPMS that had been flagged as most likely containing AI-generated content. Beyond evidence of plagiarism (paraphrasing of existing text or recycling of images without acknowledgement), they found references to nonexistent literature, references to nonexistent internal features (such as labeled theorems or formulae), and a lot of tortured phrases and meaningless gibberish. Several of the papers were found to be so similar in style and structure that they are likely traceable to the same source, presumably a paper mill that produces scholarly articles on demand and for hard dollars.$^{68}$

Sometimes a suspect paper is retracted by the publisher after being flagged on PubPeer. A typical example is a 2019 paper in the {\em Journal of Physics: Conference Series},$^{69}$ flagged on PubPeer in January 2023 for containing numerous ``tortured phrases.'' IOP Publishing retracted the paper in March 2023, citing ``concerns this article may have been created, manipulated, and/or sold by a commercial entity'' and noting a lack of ``evidence that reliable peer review was conducted on this article, despite the clear standards expected of and communicated to conference organisers.'' According to Dimensions AI, the paper was cited at least five times before it was retracted.

Based on what we have seen, the retraction of a suspect paper is still a rare event. Most of these monstrosities continue to lurk in the literature and garner citations --- by authors who use AI tools to find references relevant to their work or choose their references by title only. It will take a concerted effort by all stakeholders in the business of scientific publishing to design and implement procedures to stem the flood of fake papers and the proliferation of pseudo-science.

While Cabanac et al.\ focussed on computer science papers and many others reported on AI infection of the medical literature,$^\text{19-21,\,70-72}$ our own field has not received much attention. However, as illustrated by the afore-mentioned findings about possibly infected articles in three prominent mathematics journals published by Elsevier, we cannot assume to be immune against the virus.

To furnish an example of an infected math paper, published in a supposedly reputable math journal, we analyzed a marvelously tortured article$^{73}$ that appeared in a 2021 issue of {\em Partial Differential Equations in Applied Mathematics}, an open-access journal published by Elsevier with a clear and rigorous peer-review policy.

Supposedly a survey of spectral methods in the numerical analysis of nonlinear fluid flow problems, the article brims with tortured phrases, mangled sentences, and incomprehensible gobbledygook. Examples of tortured phrases include ``conventional differential conditions'' (ordinary differential equations?), ``liquid stream models'' (fluid flow models?), ``limited distinction techniques'' (finite-difference methods?), and ``limited volume techniques'' (finite-volume methods?). In the introduction, the authors reference a paper entitled ``{\em Highly Accurate Solutions} of the Blasius and Falkner-Skan {\em Boundary Layer Equations} via {\em Convergence Acceleration}'' (emphasis ours), creatively describing it as concerned with ``{\em profoundly precise arrangements} of Blasius and Falkner-Skan {\em limit layer conditions} through {\em combination quickening}.''

The literature referred to in the introduction of the paper is correctly referenced in the bibliography and appears to be real. Despite all the gibberish, the introduction reaps a GPT-2 output detector score of 42.81\%, indicating that it is more likely than not of human origin or has at least undergone some human editing. That said, we found numerous passages throughout the paper that are completely incomprehensible and garner GPT-2 output detector scores exceeding 99\%. The paper was processed in less than 30 days (received March~4, received in revised form April~1, accepted April~2, 2021), and it has been cited at least 18 times in subsequent publications (according to Dimensions AI).

We have been labelled alarmists for asking the question in the title of this paper; our concerns about the possible contamination of the scientific literature with AI-generated pseudo-science, fabrications, and falsities have been called ``empty fear-mongering.'' This is quite understandable. Had we been told about this cesspool four months ago, we would have been incredulous ourselves. How can it be that a professional scientist or mathematician enlists ChatGPT to write parts of a paper, copying and pasting its responses without any editing, not even eliminating telltale phrases such as ``as an AI language model'' or mistakenly copied ``regenerate response'' labels? How can it be that professional scientists or mathematicians employ automatic translation tools or paraphrasing software to write papers brimming with tortured phrases, mangled sentences, and incomprehensible gobbledygook? How is it possible for such papers to pass peer review as well as editorial scrutiny and be published in supposedly reputable journals? Well, dear colleagues, it is not only possible, but happening on a large scale. Time for a reality check!

Let's say it one more time: clear-cut policies must be put in place, and urgently so. First, to discourage abuse of generative AI tools, any use of such tools in drafting a manuscript must be formally acknowledged, specifying exactly which tools were used and for what purpose. Any deliberate violation, if detected, should result in automatic retraction. Further, employing generative AI tools in refereeing manuscripts or research proposals is unacceptable. In fact, since chatbot conversations are stored and monitored, sharing a manuscript or proposal under peer review with a chatbot constitutes a violation of confidentiality akin to posting the material on the internet. More importantly, while chatbots may be able to summarize a research paper or proposal, they cannot assess its intellectual merit or discuss its place in the literature --- this is still the province of human experts. Any peer review found to violate these principles schould be automatically rejected.

Such policies would be largely in line with those published by the Committee for Publication Ethics~\cite{COPE-Position}, the Council of Science Editors,$^{31,32}$ many major publishers,$^\text{24-28}$ the World Association of Medical Editors,$^{36}$ the International Committee of Medical Journal Editors,$^{37}$ the American Medical Society,$^{38}$ the American Physical Society,$^{33}$ the American Chemical Society,$^{34}$ the Association for Computing Machinery,$^{35}$ the National Institutes of Health~\cite{NIH}, and the Australian Research Council~\cite{ARC}. We urge SIAM, AMS, and other professional organizations in our field to follow suit; we call on NSF and other agencies that provide research funding in our field to follow the lead of NIH and ARC.

Beyond clear-cut policies as proposed above, we urgently need detailed guidelines for authors, reviewers, and editors, addressing what does or does not constitute legitimate use of generative AI in scholarly writing, reviewing, and publishing.

Given that this article will appear in the {\em Notices of the AMS}, let us consider, for example, the role of generative AI as it relates to {\em Mathematical Reviews} and MathSciNet. Is it acceptable for a MathSciNet reviewer to have ChatGPT write a review of a paper they have been assigned? Probably not. That said, is it acceptable for the reviewer to have ChatGPT summarize the paper, then evaluate the paper based on this summary rather than the full text? Maybe yes. Yes or no, the temptation to do so is almost irresistible! Why go through the trouble of reading a paper and summarizing its content, given that ChatGPT can generate a decent summary in mere seconds? Given that most MathSciNet reviews are anyway little more than summaries, should we just give up on human reviewers and let ChatGPT do the job? That would save a lot of us a lot of time and would most certainly improve the overall quality of the English! --- We are not proposing this in earnest; but we do wish to provoke some serious discussion in the community.

Clearly, there are legitimate applications of generative AI in scholarly writing. Specifically, ChatGPT can help authors improve their writing, in terms of spelling, grammar, syntax, choice of words, structure of the narrative, and overall presentation. It is also quite a capable translator, which may be useful for authors who wish or need to write in English, but lack proficiency in the language. It can read, write, and improve LaTeX code. It can summarize articles, which may be a great help when searching the literature. ChatGPT Plus, with suitable plugins enabled, can search the internet for open-access literature and help with reference management. It can also facilitate computation, data analysis, and visualization. In Section~\ref{Sensible} we describe these capabilities in more detail.

As long as sensible applications of generative AI in scholarly writing are formally acknowledged, and as long as authors take care in checking and double-checking any AI-generated content for soundness and accuracy, we don't see a problem. That said, we have encountered numerous cases, where authors used generative AI without acknowledging the fact and without as much as diligently reading the output, let alone checking for soundness and accuracy. Worse, generative AI has been used to produce pseudo-scientific papers devoid of any scholarly value. Worse yet, it has been used to conceal plagiarism: simply lift a text from someone else's work, ask ChatGPT to rewrite it in a different style, with different wording, then copy and paste the output into your own work. Unless you are sloppy enough to copy and paste also the ``regenerate response'' label, nobody might ever notice. Of course, it doesn't take artificial intelligence to generate this kind of concealed plagiarism ---  human dishonesty is all it takes. However, AI makes the job so much easier that AI-assisted plagiarism is a serious concern.

In an earlier version of this article we wrote that, unless the mathematics and wider science community takes appropriate action, ``it is not only perceivable, but unavoidable, that wide-spread use of ChatGPT and other chatbots for purposes of scholarly writing will contaminate the scientific literature with potentially serious yet hard to detect errors and fabrications, from erroneous DOIs to phantom references, allusions to imaginary scholars with fantasy biographies, and factually wrong descriptions of prior research.'' By now we know that such contamination has been occurring for some time, and on a fairly large scale. Given the universal availability of generative AI tools such as ChatGPT, the problem is presumably growing exponentially by now.

This {\em is} a watershed moment. The integrity of the scientific discourse {\em is} on the line.

\section{Experiments with ChatGPT}\label{Experiments}

About half a year ago, we decided to experiment with the free version of ChatGPT, publicly released by OpenAI on November  30, 2022. Initially, we thought of it as entertainment rather than research. We would quiz ChatGPT on questions of calculus or linear algebra. We would marvel at its abysmal performance and the fact that it reacts much like a subpar college student, with superior factual knowledge, yes, but lacking the most basic algebra and reasoning skills and making all the standard mistakes that any teacher of entry-level college mathematics has witnessed time and again.

Some of these deficiencies of ChatGPT and other chatbots will likely disappear soon enough. Already ChatGPT Plus, in the most advanced version powered by GPT-4, can call external APIs such as Wolfram Alpha,$^{74}$ a plugin that greatly enhances its algebraic and computational capabilities. A recent paper by Microsoft scientists,$^{75}$ reporting on early experiments with GPT-4, describes impressive mathematical reasoning capabilities, many persistent deficiencies not withstanding. Further improvements can be expected in the near future.

While quizzing ChatGPT on calculus is entertaining rather than disturbing, we soon discovered some very serious issues. As discussed earlier, ChatGPT and other chatbots are prone to ``hallucinations'': they routinely fabricate content and present this fictitious content authoritatively and convincingly. In the case of ChatGPT, the frequency and depth of these fabrications are nothing short of astounding.

Specifically, when questioned about research-level mathematical problems and related literature, ChatGPT typically responds in a coherent and seemingly competent manner, explaining the problem and its context, producing complete references to what appear to be relevant publications, and concisely summarizing their content. Typically, the titles and summaries of these references will closely mirror the language and wording employed by the user in their prompts (a feature technically known as ``sycophancy''): ChatGPT is talking your language and finding exactly what you are looking for! However, the user who then tries to locate the references identified by ChatGPT will soon realize that practically all of them are fake: they simply do not exist. They are fabrications, concocted to please the user. (Of course, ChatGPT is not actually capable of ``concocting content'' or ``pleasing the user''; but its algorithms are clearly designed to ensure a positive user experience.)

In \cite[Section~2]{Supplement} we illustrate these disturbing features in the context of one specific conversation with ChatGPT, which started as a genuine inquiry rather than a test. A verbatim transcript of this rather long and meandering conversation is also provided.

Subsequently, we performed more systematic experiments, asking ChatGPT to summarize research papers, or to retrieve abstracts or reviews thereof.

When provided solely with the URL or DOI or arXiv identifier of a given paper, ChatGPT will readily provide bibliographical information and a concise summary of what it claims is the paper in question; alas, usually it is not. Many times, the ``paper'' that ChatGPT references and summarizes simply does not exist; also the alleged authors may be fictional.

When provided with the full title of the paper in question, ChatGPT's summary will usually seem plausible, employing language matching the title and terminology commonly used in the field, but it may well include factually wrong statements; also bibliographical details, such as authorship and publication data, may be wrong. Even when provided with full bibliographical information about the paper in question, ChatGPT's summary, while seemingly plausible, may describe nonexistent features and fall far short of capturing the essence of the paper.

When asked to retrieve the DOI of a given paper, ChatGPT will usually oblige, but the DOI it returns is almost always wrong, either identifying a different paper or simply nonexistent. When asked to retrieve the abstract of a paper, ChatGPT will furnish the alleged abstract, usually rendered as a direct quote, enclosed in quotation marks. Alas, the quote rarely agrees with the actual abstract of the paper, neither in wording nor in substance. This behavior seems to be independent of what information the user provides about the paper in question, be it a URL, a DOI, an arXiv identifier, partial or full bibliographical details, or any combination thereof.

When asked to retrieve the MathSciNet review of a paper, ChatGPT will occasionally admit lack of access to MathSciNet; more often, however, it will oblige and produce a fake review, complete with a bogus MR number and not bearing any resemblance to the actual review.

Of course it is na\"ive and unreasonable to ask ChatGPT, which lacks real-time access to the internet, to summarize a paper available at a given link, be it publicly accessible or behind a paywall, or to retrieve the MathSciNet review of an article. It is deeply troubling, though, that ChatGPT will readily indulge such prompts and fabricate the requested summary or review, instead of informing the user that it cannot retrieve the paper in question or access MathSciNet.

Also, while ChatGPT was trained on huge amounts of text purloined from the internet, including scientific papers and books, it can obviously not be expected to have encountered paywall-protected material, including the content of most professional journals or databases such as MathSciNet (even though it will occasionally claim that OpenAI has the resources to pay, and does actually pay, for content available by subscription only). That said, the DOI registry and the arXiv are public-access repositories, and it is not unreasonable to assume that their content, as of September 2021, the cutoff date for ChatGPT's initial training, were included in its training data. If so, ChatGPT's abysmal performance in our experiments seems hard to explain.

One might suspect that ChatGPT is simply not designed to retrieve verbatim text that it encountered during training; but this is not the case. Asked, for example, to recite the poem ``The Second Coming'' by William Butler Yeats, ChatGPT will return a complete and correct quotation. Likewise, it will correctly quote the opening lines of the novel ``Cat's Cradle'' by Kurt Vonnegut or Rutger Hauer's ``Tears in Rain Monologue'' from the movie ``Blade Runner.'' It is not clear why ChatGPT would be  able to retrieve verbatim text from works of fiction, but not the abstract of a paper posted on the arXiv that it encountered during training.

In \cite[Section~3]{Supplement}, we illustrate our observations regarding ChatGPT's information retrieval capabilities (or lack thereof) in the context of a specific conversation with the chatbot; also a verbatim transcript is provided.

We note that there exist numerous AI-powered apps, such as Elicit, Scite, Scholarcy, or Humata, which are specifically designed to assist with literature searches or to summarize articles, and which may well perform better at these tasks than ChatGPT. We have not tested any of these, nor have we experimented with Microsoft's Bing (powered by OpenAI's recently released GPT-4), Google's Bard (powered by the company's own language model LaMDA), or Meta's BlenderBot (powered by the company's open-source OPT-175B language model).

\vfil

\section{Sensible Uses of ChatGPT as an Aid in Scholarly Writing}\label{Sensible}

After all the vitriol we have spilled about the pitfalls of using ChatGPT for purposes of scholarly writing, the reader may have concluded that the only way to avoid those pitfalls is to refrain from using the tool at all; however, this conclusion would be premature. As with every technology, the key is to use it judiciously: to be aware of its capabilities, but also of its limitations; to put its capabilities to work, while not asking for the impossible.

A peculiar issue with ChatGPT is that it will happily indulge most any user request, be it sensible or not. If it cannot fill your request, instead of telling you so, it will likely fudge a response, and typically, the response will sound perfectly reasonable and authoritative, even though it may be completely bogus. We have seen multiple examples of this behavior in our conversations with the chatbot~\cite[Sections~2 and~3]{Supplement}. The obvious remedy is to avoid unreasonable requests. This, however, is easier said than done. OpenAI is not very transparent about ChatGPT's capabilities and limitations, and ChatGPT itself has no idea what it can or cannot do. When quizzed about its capabilities and limitations, it tends to get entangled in contradictions.

We are aware of a few articles on good practices for scientific writing with ChatGPT.$^{76,77}$ Compiling a more comprehensive practical guide might be a worthwhile endeavor, but is clearly beyond the scope of this paper. Here, we will only briefly discuss a few capabilities of ChatGPT that are potentially useful for scholars. Some of these capabilities are accessible only via ChatGPT Plus, the subscription version of ChatGPT, which is powered by the most recent incarnation of OpenAI's large-language model, GPT-4, and can be enabled to use external APIs or plugins, including several that provide real-time access to the internet.

Most of the jobs addressed below require ChatGPT to process text it has been fed by the user. Plain text is the preferred input format, and pdf documents should be converted to plain text before input; but also LaTeX source code is acceptable and may be preferable in case of a mathematical manuscript.

There is a limit to the length of text that ChatGPT is able to process in a single prompt-and-response interaction: input and output combined should not exceed 4096 ``tokens'' (words, punctuation marks, special characters, symbols), very roughly the equivalent of three to six pages of a journal article or book. To leave enough space for the response, the input text should, in fact, be shorter than this.

Depending on the version of ChatGPT you are using, it may or may not allow you to input text that exceeds the 4096-token limit. If so, ChatGPT will process only a chunk of it. Be aware that, unless prodded, ChatGPT won't tell you that it didn't arrive at the end of the text! If you inquire, however, it will tell you whether it processed all or just part of the input. If it processed only a chunk, it may be able to go back and process additional chunks. Even when technically possible, inputting text in excess of the 4096-token limit is best avoided.

\subsection{Translating Text}\label{5.1}

ChatGPT is quite a capable translator, at least between the languages we are familiar with, English, German, Italian, and, to a lesser degree, Hebrew. According to a recent news article,$^{78}$ it is also quite proficient in Chinese, albeit somewhat less so than Baidu's recently released chatbot Earnie. While we have not performed systematic experiments, circumstantial evidence suggests that ChatGPT is better than other translation apps that we have used in the past. Obviously, this capability is potentially useful, for example, to authors who need or wish to write in English, but lack proficiency in the language. Practical advise: don't input much more than a page at a time!

\subsection{Improving Poor Writing}\label{5.2}

You can feed ChatGPT a text and ask it to suggest improvements, in terms of spelling, grammar, syntax, choice of words, structure of the narrative, and overall presentation. You can also ask it to rewrite the text on its own, in accordance with the rules of penmanship that it has been taught. This may be useful to authors who don't consider themselves distinguished wordsmiths. For advice on formatting and admissible text length, refer to the comments preceding Section~\ref{5.1}.

\subsection{Improving LaTeX Code}\label{5.3}

ChatGPT can read and write LaTeX code. You can feed it a messy LaTeX source file and ask it to suggest improvements or to rewrite and streamline the code. You can even ask it to modify the code so as to match the style of a particular publisher or journal; but the process is so cumbersome that you may prefer to do the job yourself, at least if you are using the standard version of ChatGPT. Obviously, \text{GPT-4} with internet browsing enabled is more helpful, as it can retrieve style files from publishers' websites. Practical advice: input your LaTeX code in snippets rather than large chunks!

\subsection{Summarizing Text}\label{5.4}

ChatGPT is quite adept at summarizing text that it has been fed or that it retrieved from the internet. To obtain a summary of a paper in your possession, feed ChatGPT a plain-text version (or the LaTeX source file, if available). There are a number of tools to convert pdf to text, but not all are created equal. Adobe Acrobat Reader offers a ``save as text'' option, which is convenient and works reasonably well for standard text, but is not adequate for converting displayed math or other non-textual elements of the original pdf document.

Whether you input plain text or LaTeX code, unless it's a very short paper, you will need to chop it into sensible chunks, for example, the sections or subsections of the paper. ChatGPT can summarize each chunk separately or generate a summary of the entire paper after it has processed all the chunks. You can ask for a brief (``high-level'') summary or a detailed description, including technical details or avoiding technical jargon.

If the paper is publicly available on the internet, say, on the arXiv, ChatGPT Plus can usually retrieve it, using, e.g., a plugin called ``ScholarAI.'' Unless it's a a very short paper, ScholarAI will retrieve it in chunks, suitable for processing by ChatGPT. One drawback: ScholarAI will not actually retrieve the full text of the paper, but drop all non-textual elements such as displayed math, tables, and figures.

Another plugin capable of retrieving documents from public-access direct-download links is ``Link Reader.'' Compared to ScholarAI, Link Reader is better at handling non-textual elements such as displayed math, but does not support ``retrieval in chunks'' and hence delivers usually a severely truncated version of the requested document.

In any case, you need to stay alert! Don't trust ChatGPT blindly, as it will not inform you of its limitations. For instance, when asked to summarize a paper available online, it will happily furnish a summary, based on whatever ScholarAI or Link Reader delivered, even if it was just a snippet of the actual paper. You can look at the output produced by ScholarAI or Link Reader by clicking on a link just above ChatGPT's response. However, this output is not meant for human consumption and thus very hard to read. As it was aptly put in the title of a paper we cited earlier:$^{77}$ ``ChatGPT is a Remarkable Tool --- For Experts.''

\subsection{Searching the Literature}\label{5.5}

ScholarAI can search the open-access scientific literature for abstracts matching two to six keywords, specified by the user or chosen by ChatGPT in accordance with the user's prompt. The search can be narrowed by publication years; results can be sorted by relevance or other criteria. Once a matching abstract has been found, ScholarAI is usually able to retrieve the full text of the paper or at least provide a download link. Relevant citations can be saved to a reference manager.

While all of this sounds great in theory, the practice is much less appealing. For one, it is not clear at all what sources ScholarAI draws upon, and frequently the search results seem rather random. We faked a search, having some particular references in mind and providing exactly matching keywords. ScholarAI dug up a number of loosely related works, but not the much more relevant ones we had in mind, not even after we specified the exact publication year, and not even after we provided the names of the authors. When we suggested that preprints or postprints of relevant papers might be available on the arXiv, ChatGPT invoked the Link Reader plugin, asking it to search www.arxiv.org for work by the authors whose names we had provided. Link Reader succeeded, listing the references we were after as its top search results.

Obviously, neither ScholarAI nor Link Reader can access content behind a paywall or otherwise restricted. But even sites that are freely accessible to humans may be off-limits for these apps, as the sites may be blocking ``web-crawling'' or any access attempts by bots. We suspect, for example, that the abstracts of papers published in SIAM journals, while freely available to humans, are not accessible to apps such as ScholarAI and Link Reader.

\subsection{Computation, Data Analysis, and Visualization}\label{5.6}

ChatGPT Plus can be enabled to access Wolfram Alpha$^{74}$ and, thereby, a plethora of computational, analytical, and graphical capabilities. As we have yet to perform systematic experiments, we defer a detailed discussion to a future article.

\subsection{Words of Caution}\label{5.7}

At this point, we feel the need to put a damper on the reader's possibly growing enthusiasm about all these exciting capabilities. To begin with, let's recall that plugins such as ScholarAI, Link Reader, and Wolfram Alpha, are presently still beta features, for good reasons, and things are changing on a daily basis. As of this writing, one of the latest features, ``Browse with Bing,'' is temporarily disabled, because it could apparently bypass paywalls and gain access to protected content.$^{79}$

Moreover, ChatGPT's interaction with active plugins is quite unstable, to say the least. For one, it tends to overuse them, invoking them frequently when doing so is useless or even detrimental. During a routine task, such as summarizing a given input text, in chunks, as necessitated by the 4096-token processing limit, ChatGPT may repeatedly veer off the rails, calling up ScholarAI or Link Reader for no apparent reason, then summarizing text found who knows where, but completely unrelated to the job at hand. The user needs to remain alert and vigilant, ready to intervene when necessary. ChatGPT, always your polite and well-educated assistant, may respond to such intervention by saying ``Okay, let's get back on track!'' At least it's funny$\ldots$

And, of course, at any moment ChatGPT may start hallucinating. In one experiment, we asked it to translate a short essay from English to Italian. It complied, doing a very decent job, but then stopped in mid-sentence, a few paragraphs short of the end. When asked to resume where it had left off, ChatGPT issued a continuation of the translated text that looked completely plausible, in substance as well as style. Plausibility not withstanding, this continuation was fabricated, bearing no relation to the original text. Obviously, something had gone wrong in processing, and ChatGPT seemed temporarily unable to process the final paragraphs of the original essay. Yet instead of admitting the problem, ChatGPT fudged the missing part of the translation, producing a beautiful example of an auto-complete gone rogue.

Some people say that large language models are nothing but giant auto-complete machines,$^{80}$ and there's a grain of truth to that. Maybe we shouldn't be surprised, then, to see ChatGPT hallucinate. Maybe we should be surprised that, most of the time, it doesn't\ldots

\vfil


\begin{thebibliography}{99}

\footnotesize

\bibitem{Supplement} P.~G.~Schmidt and A.~J.~Meir, Supplementary Material for This Article, SMU Scholar, Mathematics Research 9, 2023\\
\url{https://scholar.smu.edu/hum\_sci\_mathematics\_research/9}

\bibitem{Grimaldi} G.~Grimaldi and B.~Ehrler, AI et al.: Machines are About to Change Scientific Publishing Forever, ACS Energy Letters 8(1):878--880, January 4, 2023\\
\url{https://doi.org/10.1021/acsenergylett.2c02828}

\bibitem{Stokel-Walker} C.~Stokel-Walker, ChatGPT listed as author on research papers: many scientists disapprove, Nature News, Nature 613:620--621, January 18, 2023\\
\url{https://doi.org/10.1038/d41586-023-00107-z}

\bibitem{Thorp} H.~H.~Thorp, ChatGPT is fun, but not an author, Editorial, Science 379(6630):313, January 26, 2023\\
\url{https://doi.org/10.1126/science.adg7879}

\bibitem{Science} Editorial Policies, Science Journals\\
\url{https://www.science.org/content/page/science-journals-editorial-policies}

\bibitem{NatureEditorial} Tools such as ChatGPT threaten transparent science; here are our ground rules for their use, Editorial, Nature~613:612, January 24, 2023\\
\url{https://doi.org/10.1038/d41586-023-00191-1}

\bibitem{COPE-Position} Authorship and AI tools, Position Statement, Committee for Publication Ethics (COPE), February 13, 2023\\
\url{https://publicationethics.org/cope-position-statements/ai-author}

\bibitem{AMS} Use of Artificial Intelligence in Authoring, Ethical Guidelines and AMS Journal Policies, American Mathematical Society (AMS), 2023\\
\url{https://www.ams.org/publications/journals/policies/policies}

\bibitem{AMS-Advisory} Advisory Group on Artificial Intelligence and the Mathematical Community, American Mathematical Society (AMS), July 2023\\
\url{https://www.ams.org/about-us/governance/committees/ai-charge}

\bibitem{NIH} NIH Office of Extramural Research, The Use of Generative Artificial Intelligence Technologies is Prohibited for the NIH Peer Review Process, NIH Notice Number NOT-OD-23-149, June 23, 2023\\
\url{https://grants.nih.gov/grants/guide/notice-files/NOT-OD-23-149.html}

\bibitem{ARC} Policy on Use of Generative Artificial Intelligence in the ARC's Grants Programs, Version 2023.1, Australian Research Council (ARC), July 7, 2023\\
\url{https://www.arc.gov.au/about-arc/program-policies/policy-use-generative-artificial-intelligence-arcs-grant-programs}

\bibitem{Weise} K.~Weise and C.~Metz, When A.I.\ Chatbots Hallucinate, New York Times, May 1, 2023\\
\url{https://www.nytimes.com/2023/05/01/business/ai-chatbots-hallucination.html}

\bibitem{OpenAI} OpenAI, GPT-4 Technical Report, arXiv:2303.08774v3, March 27, 2023\\
\url{https://arxiv.org/abs/2303.08774v3}

\bibitem{Cabanac1} G.~Cabanac and C.~Labb\'e, Prevalence of nonsensical algorithmically generated papers in the scientific literature, Journal of the Association for Information Science and Technology 72 (12):1461-1476, May 25, 2021\\
\url{https://doi.org/10.1002/asi.24495}

\bibitem{Cabanac2} G.~Cabanac et al., Tortured phrases: A dubious writing style emerging in science: Evidence of critical issues affecting established journals, arXiv:2107.06751, July 12, 2021\\
\url{https://arxiv.org/abs/2107.06751}

\bibitem{Athaluri} S.~Athaluri et al., Exploring the Boundaries of Reality: Investigating the Phenomenon of Artificial Intelligence Hallucination in Scientific Writing Through ChatGPT References, Cureus 15(4):e37432, April 11, 2023\\
\url{https://doi.org/10.7759/cureus.37432}

\end{thebibliography}
\end{document}